\documentclass[aps,pre,amsmath,amssymb,reprint,twocolumn]{revtex4-2}
\usepackage{graphicx}
\usepackage{enumerate}
\usepackage{amsmath}
\usepackage[caption=false]{subfig}
\usepackage{hyperref}
\usepackage{float}
\usepackage{xcolor}
\usepackage[a-1b]{pdfx}
\usepackage{soul}
\newcommand{\diana}[1]{\textcolor{black}{#1}}
\newcommand{\anjan}[1]{\textcolor{black}{#1}}
\begin{document}

\title{Exploring the role of connectivity in disordered system}
\author{Anjan Daimari}
\author{Shivanee Borah}
\author{Diana Thongjaomayum}
\email{dianat@tezu.ernet.in}
\affiliation{Department of Physics, Tezpur University, Assam, India, 784028}

\date{\today}

\begin{abstract}
We study a minimal model of disordered systems, the random field Ising model (RFIM) on a generalized Petersen Graph, $GP (N,k)$, \diana{which consists of an interconnected outer and inner loop with $N$ nodes in each loop.} 
Each node in $GP(N,k)$ has coordination number $z=3$, and by varying \anjan{the parameter} $k$ different connections between the nodes in the inner loop can be obtained. The objective is to study whether different connectivity between nodes in these graphs affect the system’s response to an external field when the coordination number is fixed. This is of interest because critical behaviour is absent for $z\leq3$ on a random graph which has been solved exactly as well as on the honeycomb lattice in the context of RFIM. Using single-spin-flip Glauber dynamics at zero temperature, we compare the system’s response with the known case of $z=3$ random graph and the generalized Petersen graph for various connectivity $k$, albeit for the same $z$. Our study finds the absence of critical behaviour on $GP(N, k)$ highlighting the importance of coordination number over varying connectivity between the nodes. Additionally, we explore the case of directed $GP(N, k)$ and compare it with the undirected $GP(N, k)$ results.
\end{abstract}

\maketitle

\section{Introduction}

The nonequilibrium Random Field Ising Model (RFIM) has been extensively studied for understanding disordered spin systems, phase transitions and critical  phenomena \cite{sethna1993hysteresis,dhar1997zero,sabhapandit2000distribution,sethna2001crackling,kharwanglang,diana2016hysteresis,diana2017criteria,diana2019critical,mijatovic2021nonequilibrium}. RFIM has also been experimentally realized in a large number of real materials, such as diluted dipolar magnet, diluted antiferromagnets, single ferroelectric crystals and molecular magnets \cite{fishman1979random,hill1993magnetic,schechter2008liho,miga2009three,westphal1992diffuse,millis2010pure,xu2015barkhausen}, most recently on an artificial spin ice system \cite{bingham2021experimental}.   

The nonequilibrium zero temperature RFIM has been studied on various lattices to understand the role of dimensionality and coordination number in the system's response to slowly varying external field \cite{kharwanglang,diana2016hysteresis,diana2019critical,mijatovic2021nonequilibrium}. Studies on random graphs show critical hysteresis for $z\ge4$, whereas critical hysteresis is not observed for $z\le3$ \cite{dhar1997zero}. This shows that $z$ plays a crucial role in controlling the critical behavior of the system. In quasi-one-dimensional systems, magnetization does not exhibit jump discontinuities, suggesting that dimensionality and connectivity play a significant role in the determination of critical phemonena \cite{sabhapandit2004absence}. Similarly, studies on diluted triangular lattices report that reduced coordination leads to altered phase transition behavior, indicating that lattice structure is a major determinant of disorder induced phenomena \cite{kurbah2015nonequilibrium}.  

\begin{figure}[h]
    \centering
    \includegraphics[width=7cm, height=5cm]{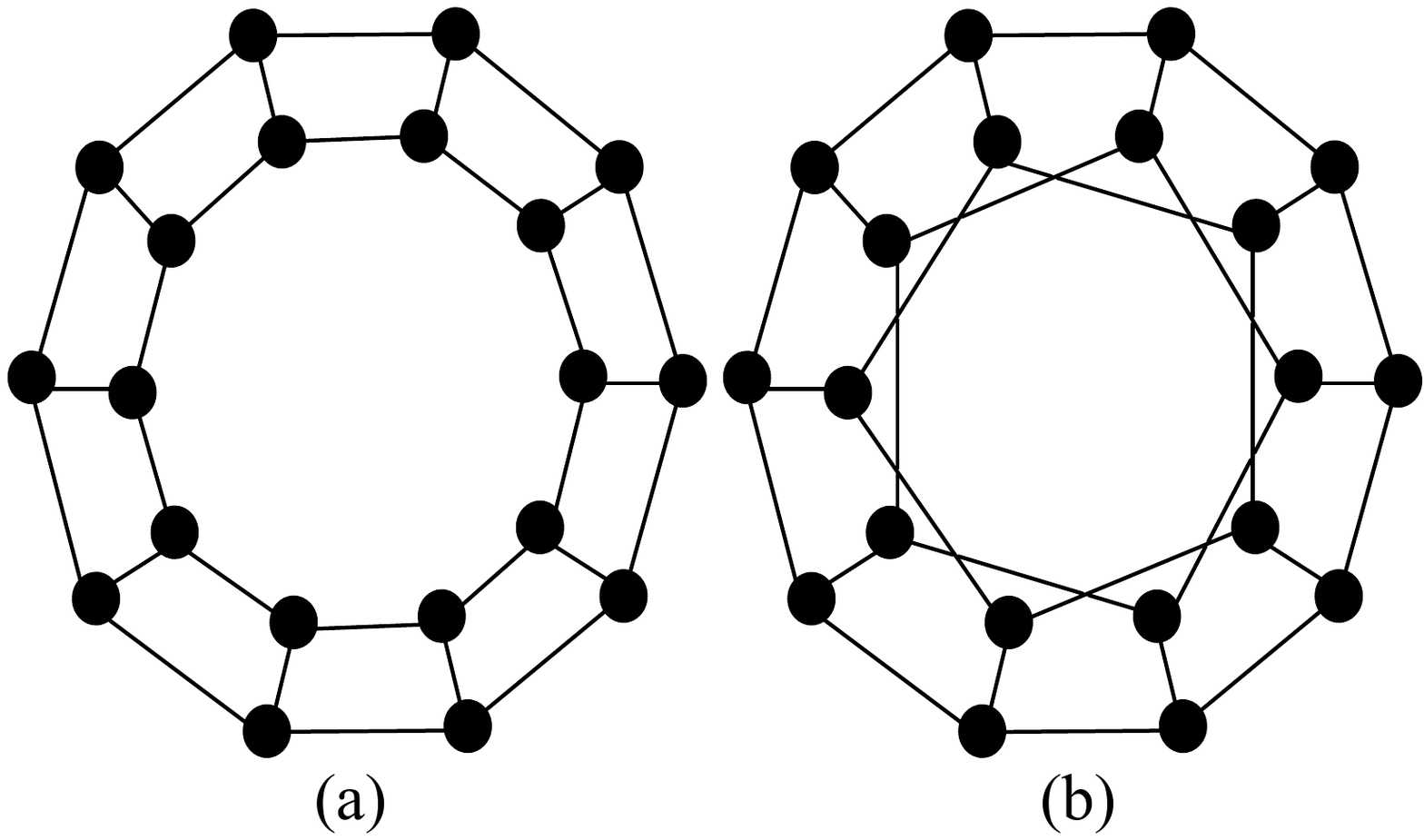}
    \caption{Schematic representation of $GP(N,k)$: (a) $GP(10,1)$  and (b)  $GP(10,2)$. $k$ determines how nodes in the inner loop are connected in $GP(N,k)$.
}
    \label{fig:i1}
\end{figure}

In this study, we apply nonequilibrium zero temperature RFIM to $GP(N,k)$. $GP(N,k)$ is a class of structured finite graphs that consists of connected inner and outer loops with $N$ nodes in each loop, first introduced by Coxeter \cite{coxeter1950self}, shown in FIG. \ref{fig:i1}. \anjan{We label the nodes of the outer loop as 1, 2, ..., $N$ and the nodes of the inner loop as $N+1$, $N+2$, ..., $2N$.} $k$ is a parameter with the condition \text{$1 \leq k < N/2$} such that any site $i$ in the inner loop has $i-k$ and $i+k$ as its two nearest neighbors apart from the fixed connection to node, $i-N$, of the outer loop whose other two neighbors are the adjacent nodes to the site: $(i-N)-1$ and $(i-N)+1$. By changing the value of $k$, different graphs can be obtained \cite{coxeter1950self,watkins1969theorem}. Since its inception, $GP(N,k)$ has gained immense interest and has been intently studied and analyzed to explore its practical applications like modeling of chemical isomerism \cite{balasubramanian2025combinatorics,mislow1970role,schwenk1989enumeration} and designing of interconnection network \cite{krishnamoorthy1987fault,ekinci2019reliability}.

By varying $k$, we can obtain different connectivity in the graph (or different graphs) with fixed $z=3$. Our main objective is to examine whether the system exhibits interesting trends in magnetization or critical phenomena such as jump discontinuities. We find that $GP(N,k)$ shows hysteresis and does not exhibit critical behavior. The magnetization curves corresponding to small values of \anjan{random field disorder strength} $\sigma$, for all $k$ spreads out. However, the magnetization curves for all values of $k$ collapse onto each other with increasing $\sigma$. Furthermore, we also study the directed version of $GP(N,k)$ \diana{and draw upon a comparative study with its undirected version.  }

\section{The Model}
We consider $GP(N,k)$, where each node ($i=1,2,3,...,2N$) is assigned an Ising spin  $S_i=\pm1$. Each spin interacts with $z$ nearest neighboring spins with ferromagnetic interaction strength $J>0$. The system experiences a uniform external field $h$ and a random field $h_i$ at each site $i$, sampled from a Gaussian distribution with mean, $\mu=0$ and standard deviation $\sigma$. The random fields are quenched throughout the system's evolution, whereas the spins $S_i$ evolve. This system corresponds to the Random Field Ising Model \cite{nattermann}, and is defined by the Hamiltonian:
\begin{equation} \tag{1}
H = -J \sum_{\langle i, j \rangle} S_i S_j - \sum_i h_i S_i - h \sum_i S_i 
\label{eq:hamiltonian}
\end{equation}
We employ single spin flip Glauber dynamics at zero temperature in discrete time steps $t$ \cite{glauber1963time}:
\[
S_i(t+1) = \text{sgn} \left[ J \sum_j S_j(t) + h_i + h \right]\tag{2}
\]
Here, the summation on the right-hand side is taken over all the $z$ nearest neighbors associated to the site $i$. The system lowers its energy iteratively until it reaches a stable fixed point {$S_i^*(h)$}, for a fixed external field $h$. Thus, {$S_i^*(h)$} for each site $i$:
\[
S_i^*(h) = \text{sgn} \left[ J \sum_j S_j^*(h) + h_i + h \right] \tag{3}
\]
The system reaches a stable fixed point when all the spins orient in the same direction as their respective local fields for the fixed $h$. We begin with an adequately large negative $h$, to set the initial configuration of all the spins \( S_i(h = -\infty) = -1 \). Next, the external field $h$ is gradually \anjan{increased until the next spin becomes unstable}. 
 At this point, the external field is fixed and subsequently the system is allowed to evolve iteratively until it converges to new fixed point by taking into the account the neigbouring spins flip through the exchange interaction. This new fixed point is termed as the magnetization per site, $m(h)$. The $m(h)$ for the total $2N$ sites is characterized by:
\[
m(h) = \frac{1}{2N} \sum_i S_i^*(h) \tag{4}
\]
\diana{The above process is repeated in the increasing field $h$ therby producing the lower hysteresis $m_l(h)$, when $h$ is ramped up from $h=-\infty$ to $+\infty$. Similarly, when $h$ is swept from $h=+\infty$ to $-\infty$, we obtain the upper hysteresis $m_u(h)$ \anjan{\cite{dhar1997zero}}.}


\[
m_u(h) = -m_l(-h) \tag{5}
\]
\section{Results}
\begin{figure}[htb]
    \centering
    \includegraphics[width=8cm, height=5cm]{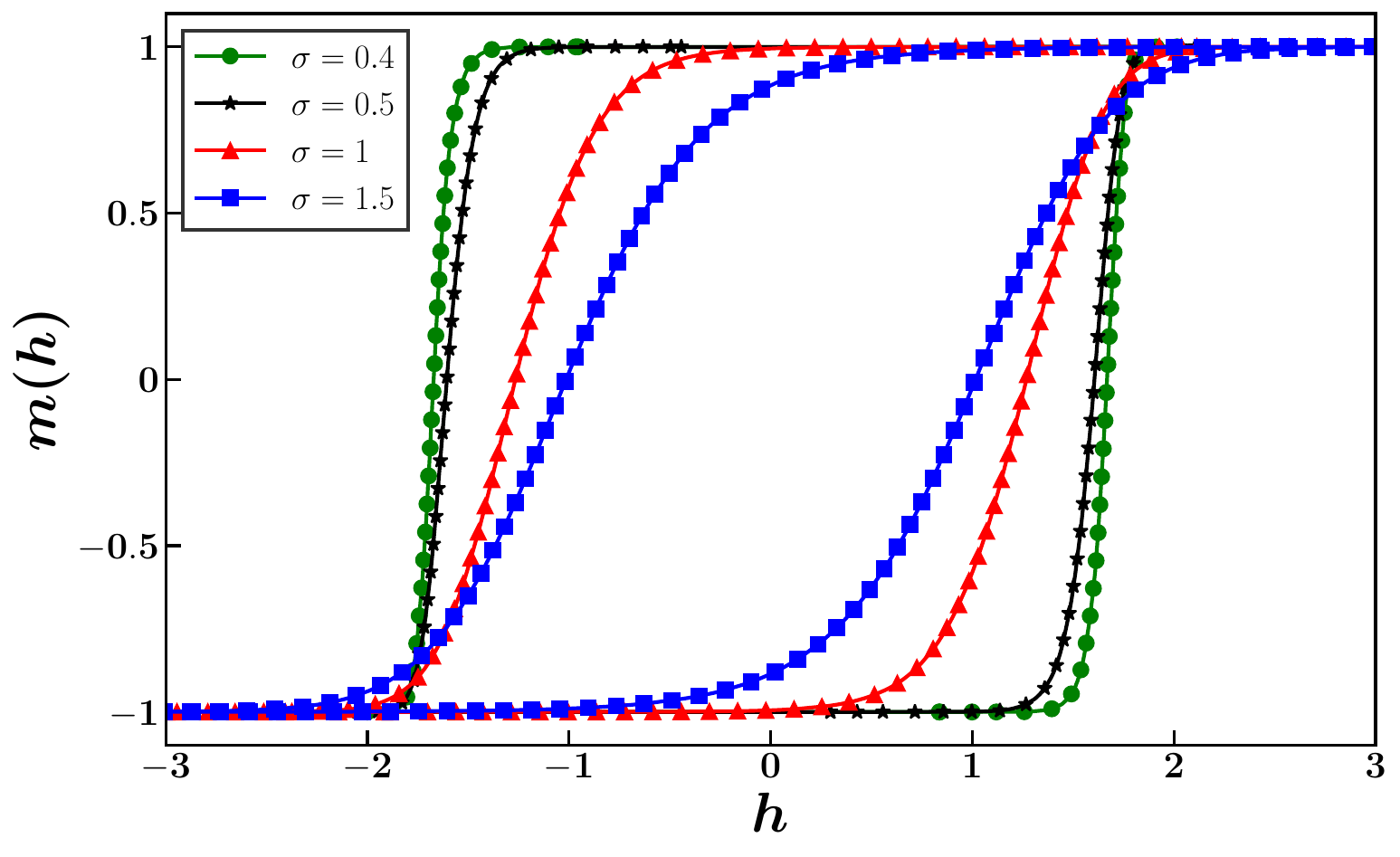}
    \caption{Hysteresis loops $m(h)$  vs $h$ on $GP(N=4999999,k=1)$ for different disorder strength $\sigma$, \anjan{obtained from single random field realization}. Solid line with circles, stars, triangles  and squares correspond to $\sigma=0.4,0.5,1$ and 1.5 respectively. The $m(h)$ curves varies smoothly for all the values of $\sigma$ and the hysteresis loops become narrower with increasing $\sigma$.} 
    \label{fig:i2}
\end{figure}
\begin{figure}[htb]
    \centering
    \includegraphics[width=8cm, height=5.5cm]{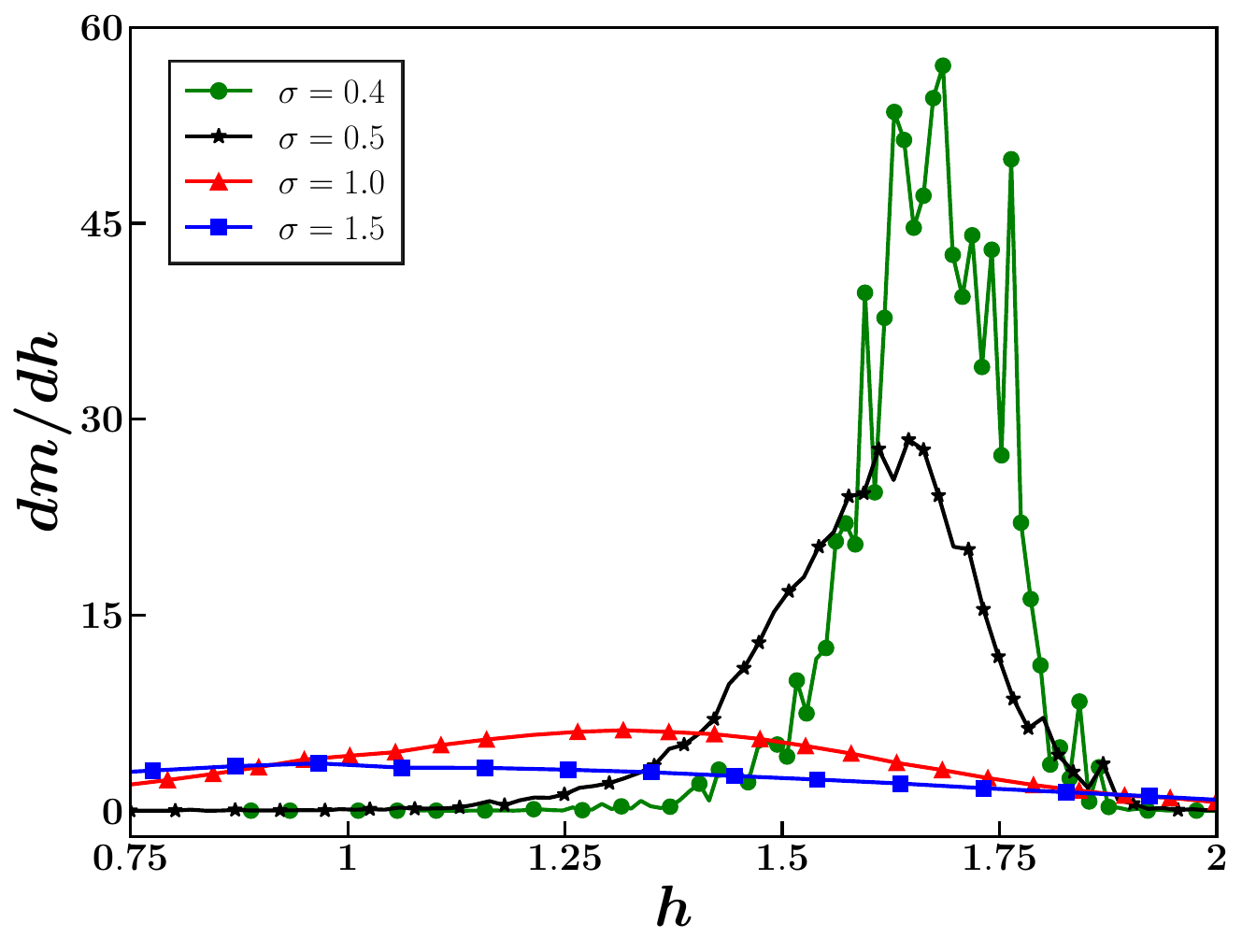}
    \caption{\anjan{Differential susceptibility, $dm/dh$ vs. external field $h$, extracted from the ascending loop of the data shown in FIG. \ref{fig:i2}. The data shown are binned over $h$. }} 
    \label{fig:i2a}
\end{figure}

FIG. \ref{fig:i2} shows the hysteresis loops of $GP(N=4999999,k=1)$ for different disorder strengths, $\sigma=0.4,0.5,1$ and 1.5 \diana{obtained for single realization of the random field}. We observe that the magnetization curves of $GP(N,k)$ for all the values of $k$ vary smoothly irrespective of $\sigma$ \anjan{although the hysteresis loops become broader and steeper with decreasing $\sigma$. Consequently, the differential susceptibility, $dm/dh$,  has broad peak over $h$ and remains within close range for the different values of $\sigma$. The response becomes smoother with broader peaks on increasing $\sigma$, see FIG. \ref{fig:i2a}.}

\begin{figure}[t]
    \centering
    \includegraphics[width=8cm, height=5.5cm]{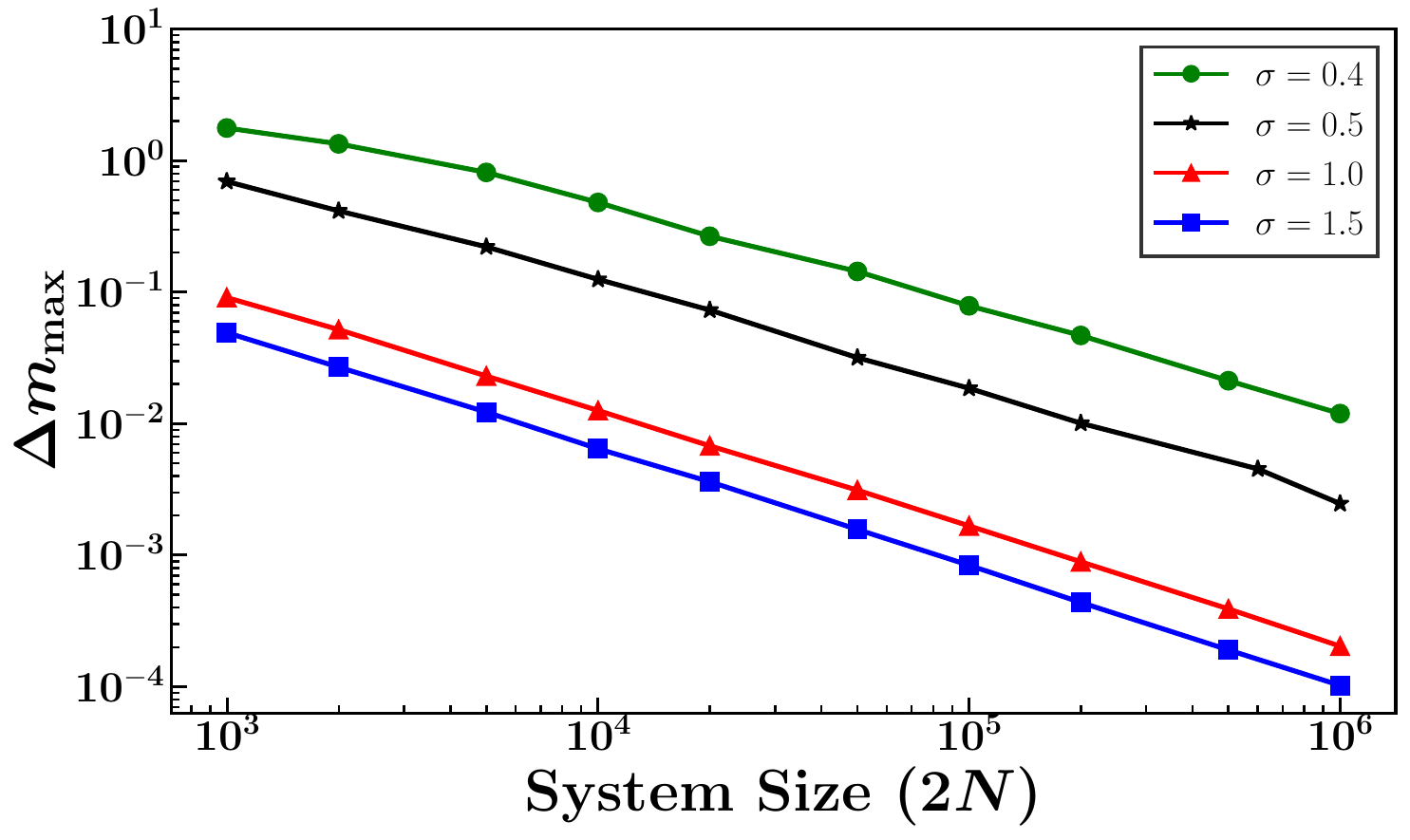}
    \caption{\anjan{Largest magnetization jump $\Delta m_{max}$ for $GP(N,1)$ as a function of system size ($2N$) for $2N$ = 998, 1998, 4998, 9998, 19998, 49998, 99998, 199998, 499998 and  999998, averaged over 100 different random field configurations.}} 
    \label{fig:i2b}
\end{figure}

\begin{figure}[htb]
    \centering
    \includegraphics[width=8cm, height=5cm]{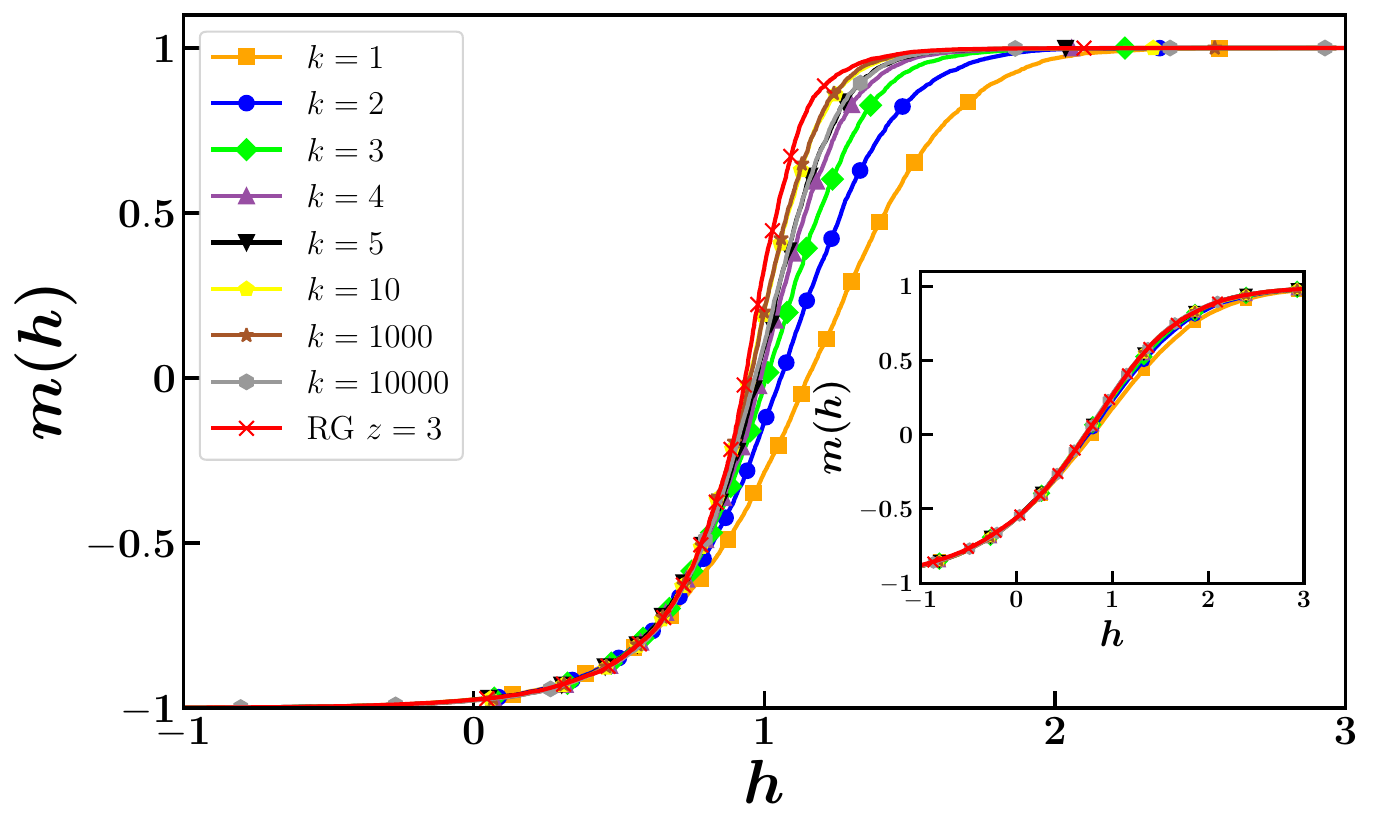}
    \caption{Plot of $m(h)$ vs $h$ on random graph with $z=3$ and $GP(N,k)$ for $k = 1, 2, 3, 4, 5, 10, 1000$ and $10000$. The main plot and inset show results for $\sigma=1.2$ and $\sigma=2.2$ respectively. The $m(h)$ curves of $GP(N,k)$ for different values of $k$ spread out for small $\sigma$ values, while $m(h)$ curves gradually approaches the $m(h)$ curve of random graph \anjan{with increasing} $\sigma$. The simulation data is obtained \anjan{for single random field realization} on $N=10^6$ for random graph and $N=499999$ for $GP(N,k)$. }
    \label{fig:i3}
\end{figure}

\anjan{Additionally, we also studied the largest magnetization jump ($\Delta m_{max}$) as a function of system size ranging from $2N=$ 998 to 999998, shown in FIG. \ref{fig:i2b}. We find that the $\Delta m_{max}$ systematically decreases with increasing system size. The above results suggest the absence of critical point in the thermodynamic limit.
The results shown are only for $k=1$. We extend the study to different values of $k$ which dictate the connectivity pattern in $GP(N,k)$.}
\diana{We observe the magnetization curves are smooth for other values of $k$ as well as shown in FIG. \ref{fig:i3}.}
Since the hysteresis behavior is symmetric, we have considered only the lower hysteresis loop.
The magnetization curves for all the values of $k$ corresponding to small values of $\sigma$ spread out.  However with increasing $\sigma$, $m(h)$ curves for all values of $k$ tend to overlap with each other as shown in the inset of FIG. \ref{fig:i3}. As the coordination number of $GP(N,k)$ is 3, we also compare our data with the results of RFIM on $z=3$ random graph. For small values of $\sigma$, the $m(h)$ curves of $GP(N,k)$ do not match the results of random graph $z=3$, shown in the main plot of FIG. \ref{fig:i3}. However, for larger values of $\sigma$ both the $m(h)$ curves of $GP(N,k)$ and  $z=3$ random graph overlap with each other. 
\begin{figure}[htb]
    \centering
    \includegraphics[width=8cm, height=5cm]{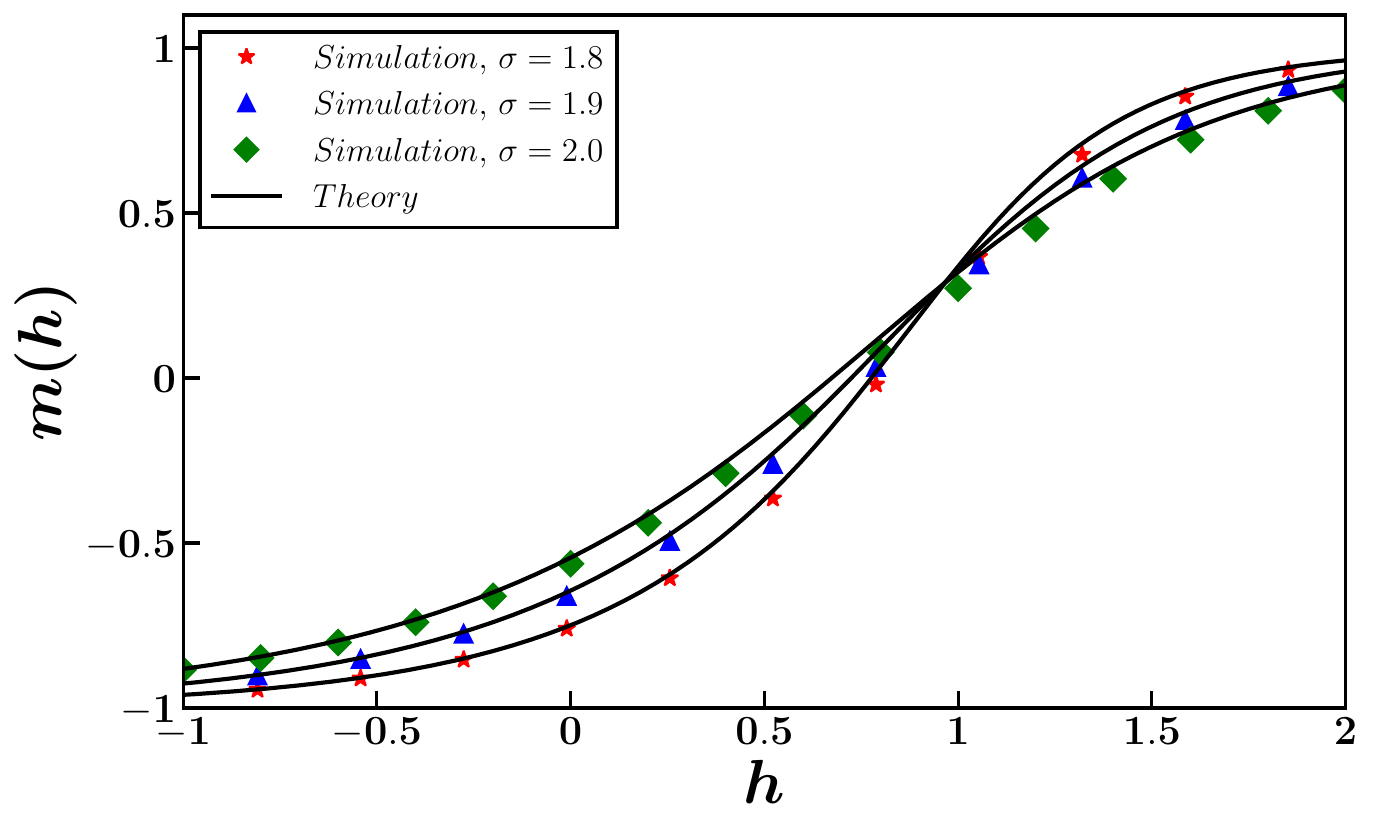}
    \caption{Comparison of $m(h)$ vs $h$ curves obtained from simulation on $GP(N=499999,k=10)$ and analytical solution for $\sigma=1.8,1.9$ and 2.2. The solid lines in the plot correspond to the theoretical predictions and the markers correspond to the simulation results. 
}
    \label{fig:i4}
\end{figure}


The analytical results on the random graph \cite{dhar1997zero} are superimposed on the numerical data of $GP(N,k)$. We consider $k=10$ because the $m(h)$ curves corresponding to $k\geq 5$ collapse onto the $m(h)$ curve of the random graph $z=3$ faster with increasing $\sigma$. 
The simulation results on $GP(N,k)$ match exactly with analytical results for higher values of $\sigma$, shown in FIG. \ref{fig:i4}.

\begin{figure}[tb]
    \centering
    \includegraphics[width=8cm, height=5cm]{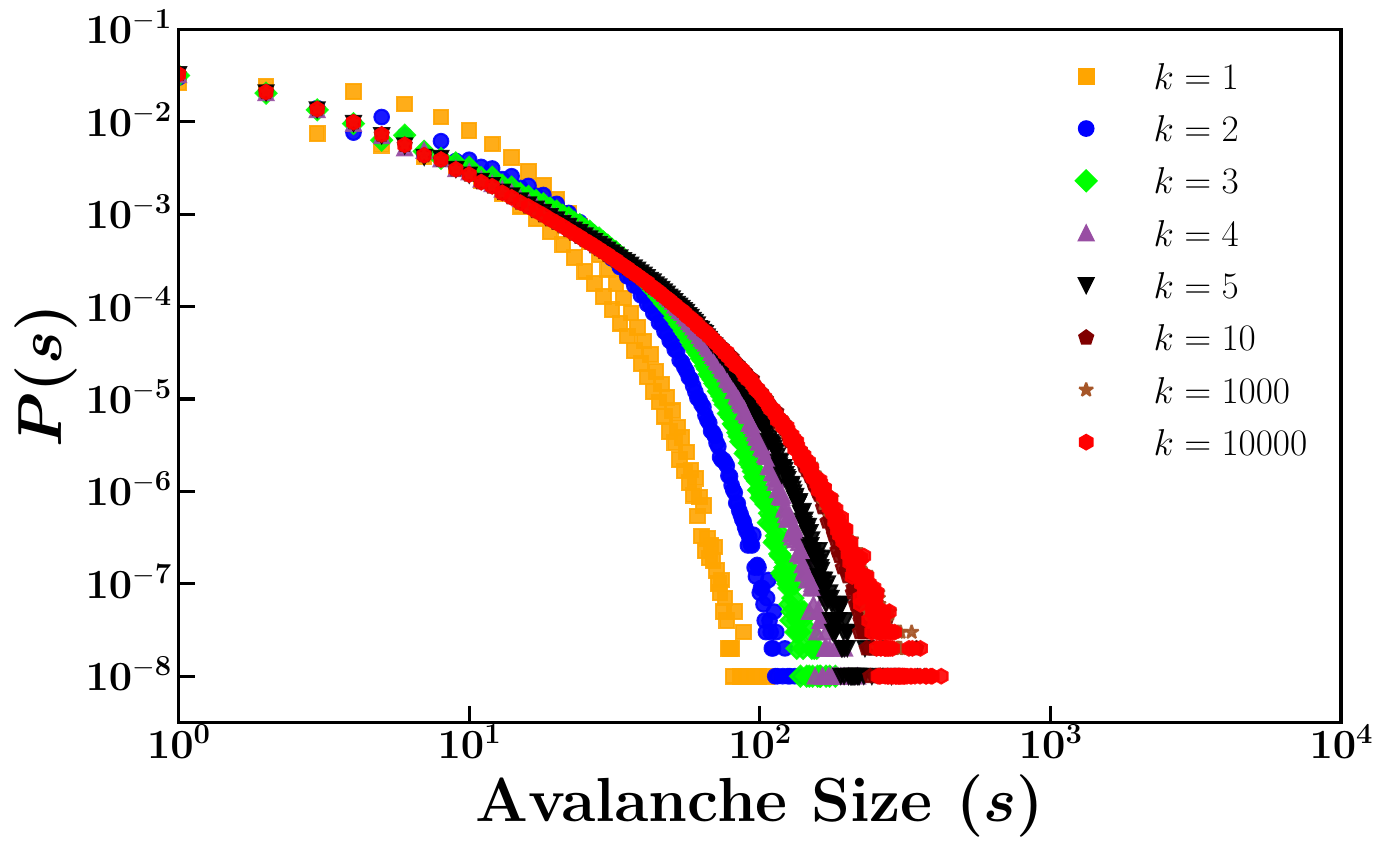}
    \caption{Log-log plot of the integrated avalanche size distribution for $k = 1, 2, 3, 4, 5, 10, 1000$ and $10000$. The data is obtained on $GP(N=499999,k)$ for $\sigma=1.2$ averaged over 100 independent configurations. The avalanche size distribution for different values of $k$ deviates from power law. 
}
    \label{fig:i5}
\end{figure}

\begin{figure}[t]
    \centering
    \includegraphics[width=7cm, height=5cm]{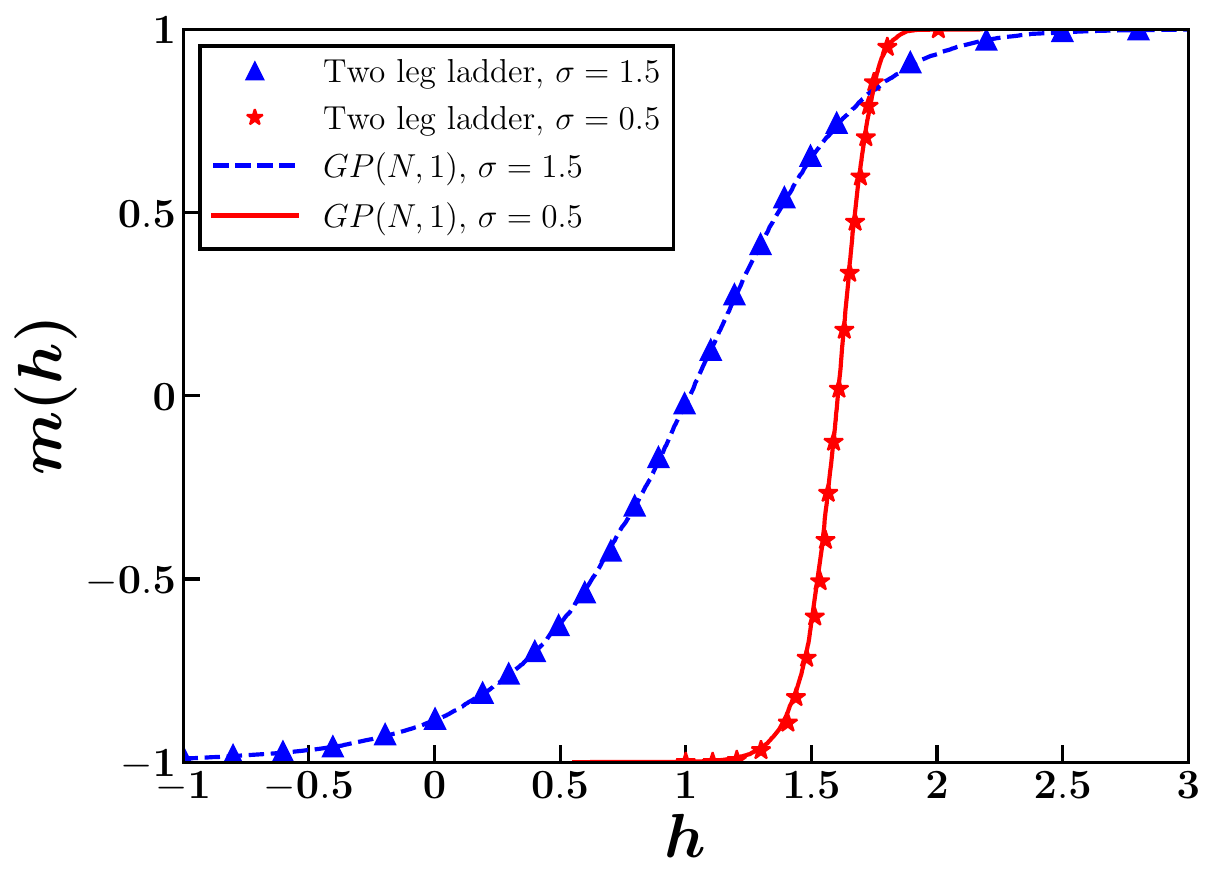}
    \caption{Magnetization curves obtained from $GP(N=499999,1)$ and the two leg ladder lattice with periodic boundary conditions. Data for two-leg ladder are from the Ref \cite{sabhapandit2004absence} correspond to $N=2^{23}$.
}
    \label{fig:Ap1}
\end{figure}

\diana{We also investigated the integrated avalanche size distribution $P(s)$ for $k=1,2,3,4,5,10,1000$ and 10000 on $GP(N=499999, k)$ averaged over 100 different random field configurations are shown in FIG. \ref{fig:i5}. The avalanche size distribution deviates from the power law and bends down spanning only few decades. This characteristic behavior supports that $GP(N,k)$ does not exhibit critical response \cite{farrow2007dynamics}.} 

The two-leg ladder with periodic \anjan{boundary} conditions \cite{sabhapandit2004absence}, and the $GP(N,1)$ has the same connectivity and lattice structure. We find that the results of two-leg ladder match perfectly with those of $GP(N,1)$ for the two values of $\sigma=0.5,$ and $ 1.5$ as shown in  
\anjan{FIG. \ref{fig:Ap1}}. \diana{The study of RFIM on two-leg ladder also reports the absence of critical behaviour for any value of $\sigma$ similar to our findings.}


\anjan{Motivated by the studies of Ref.~\cite{shukla2018hysteresis}, where introducing directed edges in the random graph has implication on the system's response,
 we also extend our study by implementing directed versions of $GP(N,k)$. First, we consider outer to inner scheme, where the nodes of the outer loop are directed towards the nodes of the inner loop. In this scheme, every nodes in the outer loop interacts with its left and right nodes in the outer loop and its corresponding node in the inner loop. However, the inner nodes interact only with the two other nodes defined by the parameter $k$ in the inner loop. Second, we consider inner to outer scheme, where we direct the nodes of the inner loop towards the nodes of the outer loop. Here, the nodes in the inner loop can interact with its inner loop nodes defined by $k$ and its corresponding node in the outer loop. However, the nodes in the outer loop interacts only with its left and right nodes.} 
\begin{figure}[h]
    \centering
    \includegraphics[width=8cm, height=5cm]{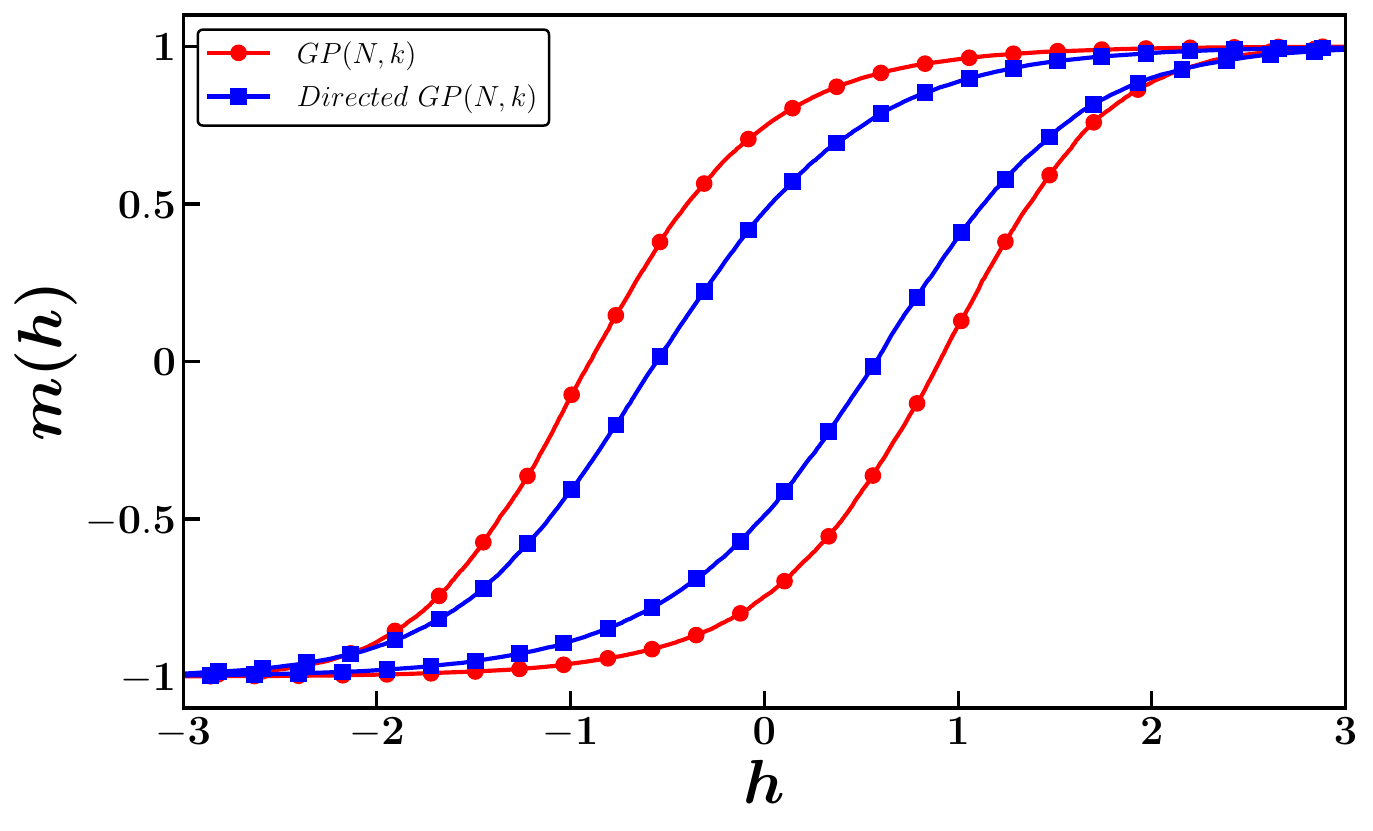}
    \caption{Comparison of hysteresis loops for undirected and directed $GP(N,k)$. The hysteresis loop of directed $G(N,k)$ shown in blue solid line with squares shows narrower hysteresis. The simulation data is obtained for $N=499999$, $k=1$ and $\sigma=1.8$. 
}
    \label{fig:i6}
\end{figure}

\anjan{The results for both the schemes of directed $GP(N,k)$ are consistent. We draw upon the comparison between the directed $GP(N,k)$ obtained from outer to inner scheme with the undirected and is shown in FIG. \ref{fig:i6}.} The hysteresis loop of directed $GP(N,k)$ is narrower compared to that undirected $GP(N,k)$. The directed $GP(N,k)$ also does not exhibit critical behavior similar to its undirected counterparts. 
\begin{figure}[htb]
    \centering
    \includegraphics[width=8cm, height=5cm]{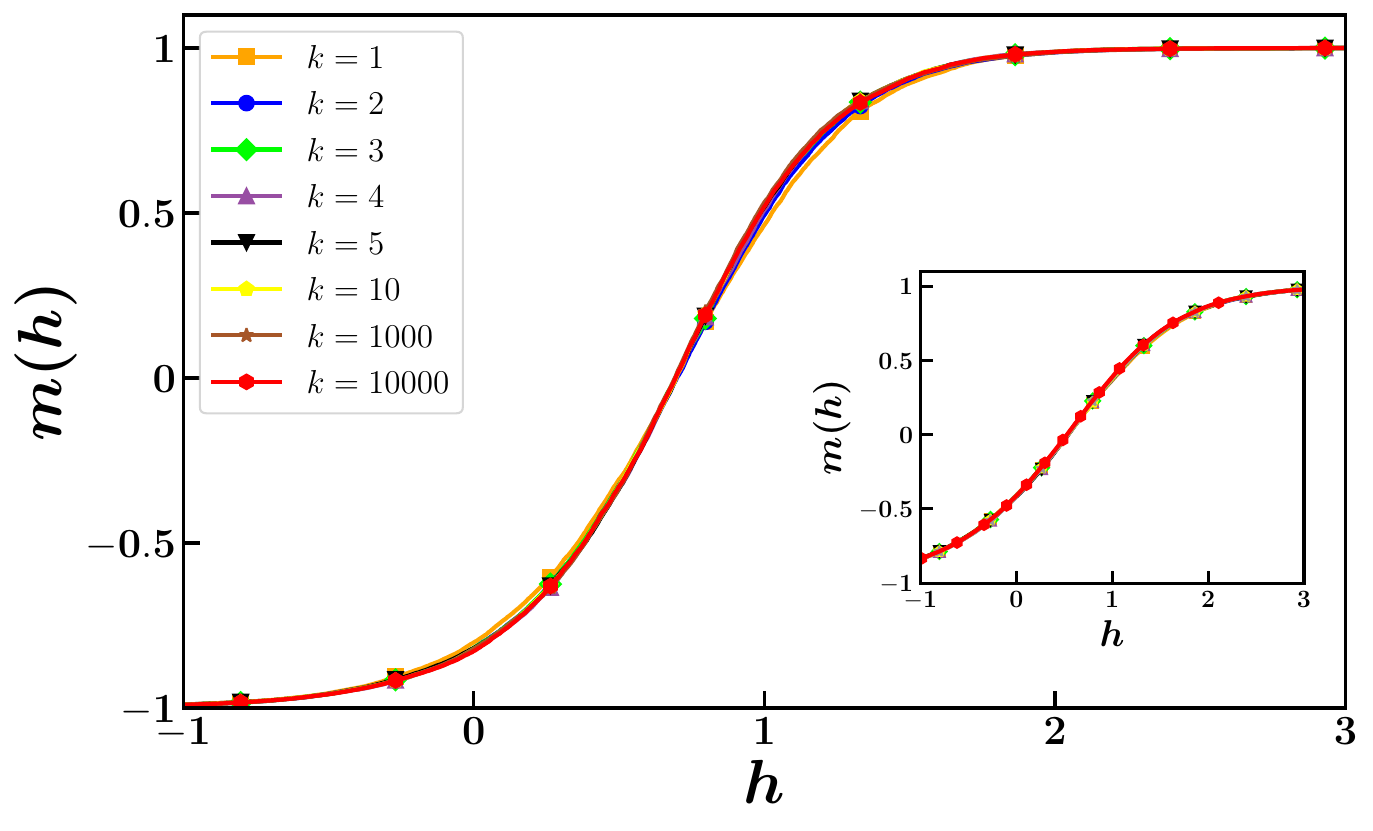}
    \caption{Plot of $m(h)$ vs. $h$ on the directed $GP(N=499999,k)$ for $k=1,2,3,4,5,10,1000$ and 10000. The main plot and inset show results for $\sigma=1.2$ and $\sigma=2$ respectively. The $m(h)$ curves for different values of $k$ do not completely overlap for small values of $\sigma$, however with increasing $\sigma$, the curves tend to align perfectly onto each other.
}
    \label{fig:i7}
\end{figure}

\begin{figure}[hbt]
    \centering
    \includegraphics[width=8cm, height=5cm]{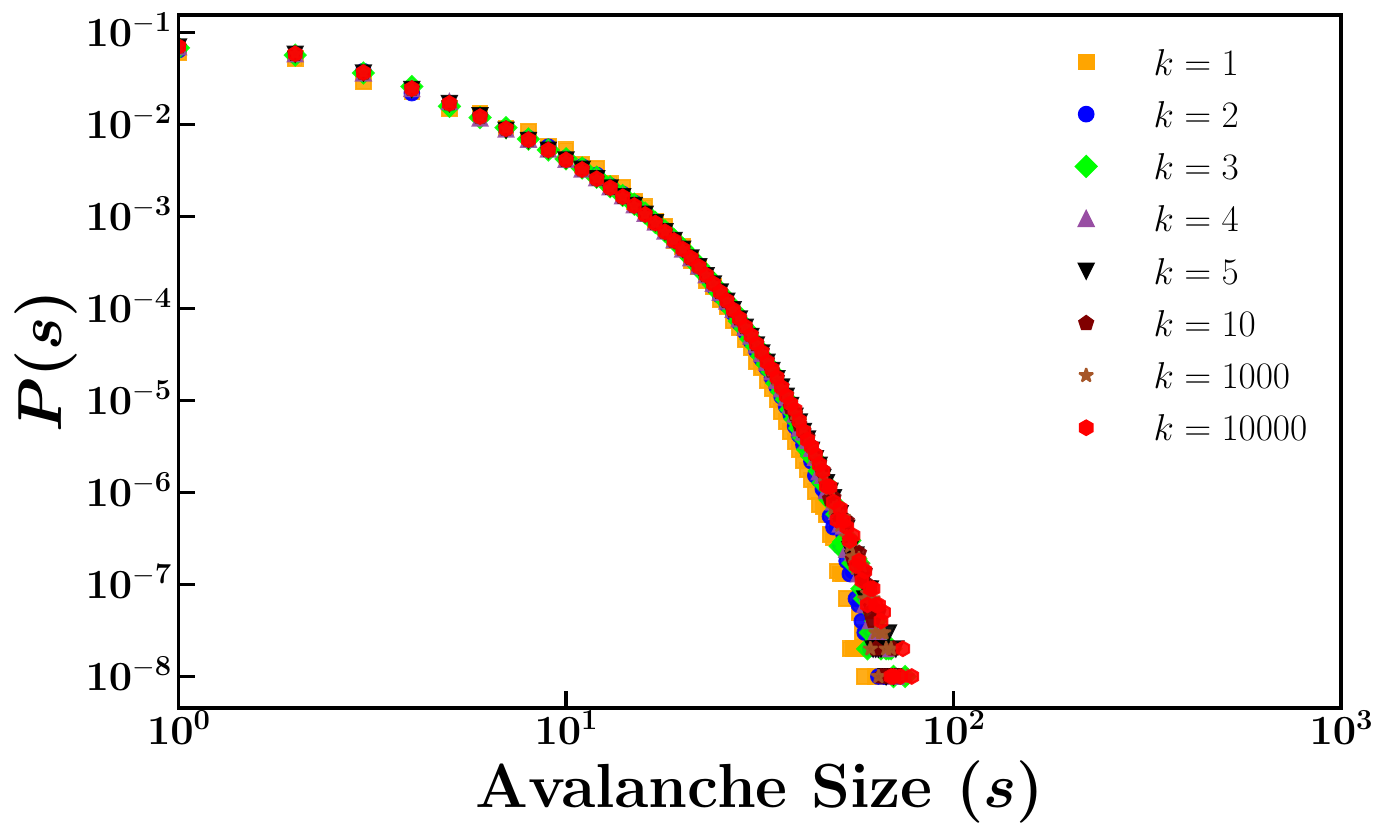}
    \caption{Log-log plots of integrated avalanche size distribution for $k = 1, 2, 3, 4, 5, 10, 1000$ and $10000$. The data is obtained on directed $GP(N=499999,k)$ for $\sigma=1.2$ averaged over 100 independent configurations. Similar to the undirected version of $GP(N,k)$, in the directed $GP(N,k)$ the avalanche size distribution for different values of $k$ shows bending down and does not span for many decades.
}
    \label{fig:i8}
\end{figure}

Further, the $m(h)$ curves of directed $GP(N,k)$ corresponding to small values of $\sigma$ for different values of $k$ do not fully overlap onto each other as shown in the main plot of FIG. \ref{fig:i7}. However, on increasing $\sigma$, the $m(h)$ curves collapses on each other, shown in the inset of FIG. \ref{fig:i7}. Similar to $GP(N,k)$ the integrated avalanche size distribution of directed $GP(N,k)$ deviates from power law behavior as evident from FIG. \ref{fig:i8}.

\section{Conclusion}
In summary, we have presented the numerical results of the zero temperature RFIM on $GP(N,k)$. We find that $GP(N,k)$ and directed $GP(N,k)$ show hysteresis but do not exhibit critical behavior, which is consistent with the finding on $z=3$ random graph \cite{dhar1997zero}. The $m(h)$ curves of $GP(N,k)$ corresponding to small values of disorder $\sigma$ are spread out irrespective of the $k$ value. However, on increasing $\sigma$ the curves merge onto one other.
The same behaviour is observed in the case of directed $GP(N,k)$.
The results of $GP(N,k)$ are consistent with the numerical and analytical results of $z=3$ random graph for higher values of $\sigma$. This ensures the availability of exact solution on the $GP(N,k)$ in the limit of large disorder.
The avalanche size distribution for both the $GP(N,k)$ and directed $GP(N,k)$ does not span for many decades and bends faster which is another signature of system showing no critical response \cite{farrow2007dynamics}.

\anjan{The absence of criticality can be understood from the underlying structure and the coordination number of the $GP(N,k)$. The key characteristics of the $GP(N,k)$ is its girth or the length of shortest closed loop. The value of girth in the $GP(N,k)$ ranges from 3 to a maximum of 8 \cite{morris2022most} depending upon the value of $N$ and $k$. 
The low girth number confines the propagation of the avalanche to farther nodes.
For a fixed $k$, increasing the total nodes $N$ has no effect  unlike the random graphs where with increasing $N$, the loop becomes larger. 
Additionally, the spectral gap of the $GP(N,k)$ becomes small as $N\to \infty$, which results in poor expansion properties \cite{dudek2016spectrum}. 
Thus, the low girth and poor expansion properties contribute to the restriction of the spins in a closed and small loop.
Increasing $k$ has some role in influencing the interaction between the distant nodes and consequently deviation in the $m(h)$ is observed in low disorder regime (FIG. \ref{fig:i3}). Nonetheless, the small loop structure in the system persists even on increasing $k$.
In case of the directed $GP(N,k)$, introduction of directed edges alters the structure of $GP(N,k)$ resulting in a mixed graph of coordination number $3$ and $2$. The cascading effect of spin flip thus gets fractured, causing narrower hysteresis loop and smaller avalanches compared to the undirected version, as evident from  FIG. \ref{fig:i6} and FIG. \ref{fig:i8}. Also, in the outer to inner directed scheme, any node $i$ in the inner loop interacts only with its two neighbors at site $i+k$ and $i-k$. Any further propagation of avalanche would be restricted within the few connected nodes in the inner loop only. Similarly, for the inner to outer directed case, the progress of the avalanche is  confined to the outer loop only. Consequently, $k$ plays a minor role in the directed $GP(N,k)$ as observed in Fig. \ref{fig:i7}.}

\anjan{$GP(N,k)$ is not a general ensemble since for mean field graphs with $z$ fixed, the loop length will increase with $N$ \cite{2ndRefereeNote}. Whereas in $GP(N,k)$, the girth length is capped at 8 even for large  $N$ and $k$. } Nonetheless, the importance of coordination number rather than the connectivity, is highlighted in the present study. The propagation of large avalanche requires a minimum connectivity to facilitate the critical response. Hence, graphs or regular lattices with $z>3$ show critical behaviour in the context of nonequilibrium RFIM. 
The study of graphs is crucial not only from the mathematical aspect, but also from the perspective of network studies.

\section*{Acknowledgements}
A.D. and D.T. acknowledge the Department of Science
and Technology, Anusandhan National Research Foundation, ANRF (then DST-SERB), Government of India, for financial support under POWER Grant No. SPG/2022/000678.
D.T. also acknowledges partial support from ANRF under the
Partnerships for Accelerated Innovation and Research (PAIR)
for the project JNU-PAIR network on Science for Sustainable
Future, File No. ANRF/PAIR/2025/000029/PAIR.

\section*{Data Availability}
The data that support the findings of this article are not
publicly available. The data are available from the authors
upon reasonable.

\bibliography{prfim}
\bibliographystyle{apsrev4-2}
\end{document}